\documentclass[superscriptaddress,aps,prl,twocolumn]{revtex4-1}
\usepackage{graphicx}
\usepackage{dcolumn}
\begin{document}

\title{Half-lives and branchings for $\beta$-delayed neutron emission for neutron rich 
Co-Cu isotopes in the r-process}

\author{P. Hosmer}
\affiliation{National Superconducting Cyclotron Laboratory, Michigan State University,
East Lansing, MI 48824, USA}
\affiliation{Dept. of Physics and Astronomy, Michigan State University, East Lansing,  MI 48824, USA}

\author{H. Schatz}
\affiliation{National Superconducting Cyclotron Laboratory, Michigan State University,
East Lansing, MI 48824, USA}
\affiliation{Dept. of Physics and Astronomy, Michigan State University, East Lansing,  MI 48824, USA}
\affiliation{Joint Institute for Nuclear Astrophysics, Michigan State University, East Lansing,  MI 48824, USA}

\author{A. Aprahamian}
\affiliation{Dept. of Physics and Joint Institute for Nuclear Astrophysics, University of Notre Dame, Notre Dame, IN 46556-5670, USA}

\author{O. Arndt}
\affiliation{Institut f{\"u}r Kernchemie, Universit{\"a}t Mainz, Fritz-Strassmann Weg 2,
D-55128 Mainz, Germany}
\affiliation{HGF Virtuelles Institut f{\"u}r Kernstruktur und Nukleare Astrophysik}

\author{R. R. C. Clement}
\affiliation{National Superconducting Cyclotron Laboratory, Michigan State University,
East Lansing, MI 48824, USA}
\affiliation{Current affiliation: Applied Physics Division, Los Alamos National Laboratory, Los Alamos, NM87545, USA}

\author{A. Estrade}
\affiliation{National Superconducting Cyclotron Laboratory, Michigan State University,
East Lansing, MI 48824, USA}
\affiliation{Dept. of Physics and Astronomy, Michigan State University, East Lansing,  MI 48824, USA}

\author{K. Farouqi}
\affiliation{Dept. of Astronomy and Astrophysics and Joint Institute for Nuclear Astrophysics, University of Chicago, Chicago, IL 60637, USA}
\affiliation{Current affiliation: Landessternwarte,
ÊÊÊÊÊUniversitŠt Heidelberg, Kšnigstuhl 12, D-69117 Heidelberg, Germany}

\author{K.-L. Kratz}
\affiliation{Max-Planck-Institut f{\"u}r Chemie (Otto-Hahn-Institut, J.-J.-Becherweg 27,
D-55128 Mainz, Germany}
\affiliation{HGF Virtuelles Institut f{\"u}r Kernstruktur und Nukleare Astrophysik}

\author{S. N. Liddick}
\affiliation{National Superconducting Cyclotron Laboratory, Michigan State University,
East Lansing, MI 48824, USA}
\affiliation{Dept. of Chemistry, Michigan State University, East Lansing, MI 48824, USA}

\author{A. F. Lisetskiy}
 \affiliation{Department of Physics, University of Arizona, Tucson, AZ 85721, USA} 

\author{P. F. Mantica}
\affiliation{National Superconducting Cyclotron Laboratory, Michigan State University,
East Lansing, MI 48824, USA}
\affiliation{Dept. of Chemistry, Michigan State University, East Lansing, MI 48824, USA}

\author{P. M{\"o}ller}
\affiliation{Theoretical Division, Los Alamos National Laboratory, Los Alamos, New Mexico 87545, USA}

\author{W. F. Mueller}
\affiliation{National Superconducting Cyclotron Laboratory, Michigan State University,
East Lansing, MI 48824, USA}

\author{F. Montes}
\affiliation{National Superconducting Cyclotron Laboratory, Michigan State University,
East Lansing, MI 48824, USA}
\affiliation{Dept. of Physics and Astronomy, Michigan State University, East Lansing,  MI 48824, USA}

\author{A.C. Morton}
\affiliation{National Superconducting Cyclotron Laboratory, Michigan State University,
East Lansing, MI 48824, USA}
\affiliation{Current affiliation: TRIUMF, 4004 Wesbrook Mall, Vancouver, BC V6T 2A3 Canada}

\author{M. Ouellette}
\author{E. Pellegrini}
\affiliation{National Superconducting Cyclotron Laboratory, Michigan State University,
East Lansing, MI 48824, USA}
\affiliation{Dept. of Physics and Astronomy, Michigan State University, East Lansing,  MI 48824, USA}

\author{J. Pereira}
\affiliation{National Superconducting Cyclotron Laboratory, Michigan State University,
East Lansing, MI 48824, USA}

\author{B. Pfeiffer}
\affiliation{Institut f{\"u}r Kernchemie, Universit{\"a}t Mainz, Fritz-Strassmann Weg 2,
D-55128 Mainz, Germany}
\affiliation{HGF Virtuelles Institut f{\"u}r Kernstruktur und Nukleare Astrophysik}

\author{ P.~Reeder}
\affiliation{Pacific Northwest National Laboratory, MS P8-50, P.O. Box 999, Richland, WA 99352, USA}

\author{P. Santi}
\affiliation{National Superconducting Cyclotron Laboratory, Michigan State University,
East Lansing, MI 48824, USA}
\affiliation{Current affiliation: Los Alamos National Laboratory, TA 35 Bldg. 2 Room C-160, USA}

\author{M. Steiner}
\author{A. Stolz}
\affiliation{National Superconducting Cyclotron Laboratory, Michigan State University,
East Lansing, MI 48824, USA}

\author{B. E. Tomlin}
\affiliation{National Superconducting Cyclotron Laboratory, Michigan State University,
East Lansing, MI 48824, USA}
\affiliation{Dept. of Chemistry, Michigan State University, East Lansing, MI 48824, USA}

\author{W. B. Walters}
\affiliation{Dept. of Chemistry and Biochemistry, University of Maryland, College Park, MD 20742, USA}

\author{A. W{\"o}hr}
\affiliation{Dept. of Physics, University of Notre Dame, Notre Dame, IN 46556-5670, USA}

\begin{abstract}
The $\beta$-decays of very neutron rich nuclides in the Co-Zn region 
were studied experimentally at the National Superconducting Cyclotron Laboratory
using the NSCL $\beta$-counting station in conjunction with the neutron detector NERO. 
We measured the branchings for $\beta$-delayed neutron emission ($P_n$ values) for 
$^{74}$Co (18$\pm$15\%), and $^{75-77}$Ni (10$\pm2.8$\%, 14$\pm$3.6\%, and 
30$\pm$24\%, respectively) for the first time, and remeasured  the $P_n$ values of 
$^{77-79}$Cu, $^{79,81}$Zn, and $^{82}$Ga. For $^{77-79}$Cu and for $^{81}$Zn we obtain significantly larger $P_n$ values compared to previous work. While the new half-lives for 
the Ni isotopes from this experiment had been reported before, we present here in addition 
the first half-life measurements of $^{75}$Co (30$\pm$11~ms) and $^{80}$Cu (170$^{+110}_{-50}$~ms). Our results are compared with theoretical predictions, and their impact on 
various types of models for the astrophysical rapid neutron capture process (r-process) is explored. We find that with our new data the classical r-process model is better able to reproduce 
the $A=78-80$ abundance pattern inferred from the solar abundances. The new data also 
influence r-process models based on the neutrino driven high entropy winds in core collapse supernovae. 
\end{abstract}


\maketitle

\section{Introduction}

The rapid neutron capture process (r-process) is traditionally believed to produce roughly half of the heavy elements beyond 
the iron region \cite{BBFH,CTT91,AGT07}. The site of the r-process is still not known with certainty. The necessary very high densities of free neutrons require extreme conditions that have been proposed to be encountered
in various sites within core collapse supernovae, for example the neutrino driven wind 
in delayed explosion models \cite{WWH92,TWJ94,BLD06}, 
jets \cite{Cam01}, fallback material \cite{FFA06}, or, maybe, prompt explosions \cite{STM01,WTI03}. Alternatively, the r-process could also 
occur in neutron star mergers \cite{RLT99}, $\gamma$-ray bursts \cite{PWH03,SuM05}, 
or quark novae \cite{JMO06}. What is known from observations 
is the pattern of isotopic and elemental abundances that the r-process produces. This r-process
abundance pattern can be extracted from the solar system abundances by subtracting the contributions
from the s- and p-processes.  It can also be observed directly in a specific class of extremely 
metal poor but r-process element enhanced stars in the halo of the Galaxy
(for a recent review see \cite{SCG08}). More than two dozen of such stars
have been found so far revealing a consistent r-process pattern for elements from Ba to the Pt peak,
but showing variations for lighter and heavier elements. 
Ongoing large scale surveys of Galactic halo stars
together with high resolution spectroscopic followups are expected to 
find many more such stars in the future. This continuously increasing body of observational 
information needs to be compared and interpreted with r-process models for the various 
sites that have been proposed. This requires a solid understanding of the 
underlying nuclear physics, that can have as much influence on the r-process abundances
as the astrophysical environment \cite{KBT93,Pfe01,Kra07}. 
Currently, nuclear physics uncertainties prevent the reliable
extraction of site specific signatures from observational data and reliable calculation
of the nucleosynthesis products of a specific r-process scenario.

Most models assume that the r-process is a sequence of rapid neutron captures and 
$\beta$-decays. Neutrino interactions and fission processes might play some role depending
on the specific r-process environment. 
Among the most important nuclear physics quantities needed in r-process models are
the $\beta$-decay half-lives of the r-process waiting points, which determine directly the 
process timescale and the produced abundance level at their location in the path.
Branchings for $\beta$-delayed 
neutron emission are also important \cite{KBT93} during and after the r-process freezes out and
the unstable nuclei along the r-process path decay towards stability. $\beta$-delayed 
neutron emissions during that stage modify the 
final abundances and increase the neutron abundance during freezeout. 
In recent years great progress has been made in experimentally determining
$\beta$-decay properties of nuclei relevant for the r-process
\cite{KPT00,SBF02,Pfeiffer02,SDA03,DKW03,KPA05,Hos05,MEH06,AHH09,PHA09}. 
Nevertheless, only a very small fraction of the r-process isotopes have been reached experimentally 
so far, most of them located near or in between the $N=50$ and $N=82$ shell closures. R-process
model calculations therefore rely on global theoretical models for the prediction 
of $\beta$-decay properties far from stability. These models have to be tested
systematically by comparison with experimental data along isotopic chains far from 
stability. 

This collaboration has reported half-life measurements of 
the waiting point nucleus $^{78}$Ni and other neutron rich Ni isotopes 
performed at the National Superconducting Cyclotron Laboratory (NSCL) at 
Michigan State University \cite{Hos05}. With this measurement, the half-lives of all the relevant 
$N=50$ waiting points in the r-process, $^{78}$Ni, $^{79}$Cu, and $^{80}$Zn, are known 
experimentally. This mass region plays a critical role in the subset of r-process models
(for example \cite{KBT93,TSK01,WTI03}) 
that are characterized by a neutron capture flow through $N=50$, where it 
represents the first major bottle-neck for the production of heavier nuclei after 
the $A=8$ stability gap.

In this paper we present $\beta$-decay half-lives and branchings for 
$\beta$-delayed neutron emission for a range of very neutron rich 
Co, Ni, Cu, and Zn isotopes obtained in the same experiment. These data provide
tests for theoretical models used to predict 
$\beta$-decay properties for r-process model calculations. In particular,
the combined measurement of the half-life and the branching for 
$\beta$-delayed neutron emission provides a stringent test
probing the $\beta$-strength function at low excitation energies and
just above the neutron threshold \cite{Kra84}. In addition, improved 
data on the branchings for $\beta$-delayed neutron emission of 
$^{78}$Cu and $^{79}$Cu are direct input in r-process model 
calculations. They determine the final abundance pattern produced by the decay 
of the abundances accumulated during the r-process at the $N=50$ waiting point nuclei. 

The region around $^{78}$Ni is also of considerable interest for nuclear 
physics. In general, doubly magic nuclei provide a testing ground 
for single particle structure and shell models, which in the
cases of $^{78}$Ni and $^{132}$Sn are uniquely located at extreme neutron excess.
Consequently, a great deal of experimental \cite{GBB98,MBF00,FHK01,SGM03,SMG04,SPG04,MGB05,VDG05} and theoretical
\cite{VDG05, EBD99,DGL00,GGF02,MPK03,Lam03,SDV04,Bor05,LBH04,LBH05} 
activity has been devoted to nuclei near $^{78}$Ni. An example
is the existence of 8$^+$ seniority isomers that is now established
in $^{70}$Ni \cite{GBB98}, $^{76}$Ni \cite{SPG04,MGB05}, and $^{78}$Zn \cite{DGL00} and is interpreted as 
evidence of the persistence of strong shell gaps out to $^{78}$Ni. 
The non-existence of these isomers in $^{72}$Ni and $^{74}$Ni \cite{SGM03} is 
now explained with a subtle change in the residual nucleon nucleon
interaction. Another example is the shift of the $f_{5/2}$ proton 
orbital observed in the odd $A$ Cu isotopes with increased filling
of the neutron $g_{9/2}$ orbital due to the monopole term in the proton-neutron
residual interaction \cite{FHK01,SDV04}. An understanding of these nuclear 
structure effects around $^{78}$Ni requires reliable shell model 
calculations \cite{DGL00,GGF02,SDV04,VDG05}. Recently some efforts have 
been made to develop a new effective shell model interaction for this region
(JJ4A)  \cite{LBH04}.  Our new data provide a test for how well these shell model calculations
extrapolate to the most neutron rich nuclei. 

\section{Setup}
  
\begin{figure*}[t]
\includegraphics[width=5cm,bb=300 220 600 500]{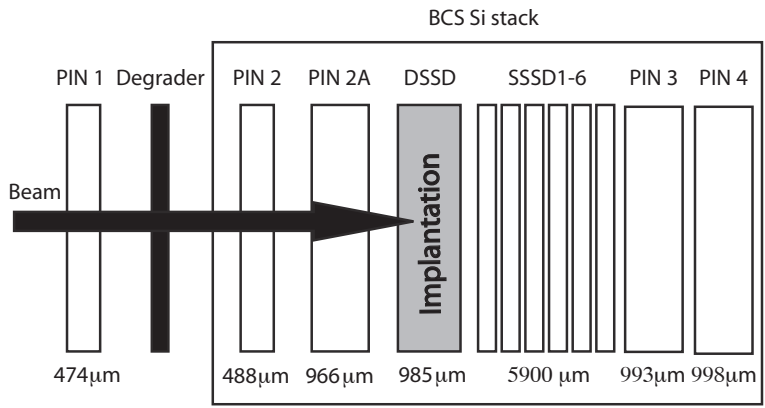}
\caption{\label{FigSetup} Schematic arrangement of the Si detectors comprising the 
NSCL Beta Counting System in the configuration used for this experiment. The degrader was
adjusted to ensure implantation within the DSSD.}
\end{figure*}

Extremely neutron rich nuclei were produced at the NSCL by fragmentation of 
a 15 pnA 140 MeV/nucleon $^{86}$Kr$^{34+}$ primary beam on a 376 mg/cm$^2$ Be target. 
After separation with the A1900 fragment separator using the $B\rho - \Delta E - B\rho$
method \cite{MSS03}, the mixed beam was implanted continuously into the NSCL Beta Counting System
\cite{Pri03} consisting of a stack of Si PIN detectors, a double-sided Si strip detector (DSSD) 
for implantation of the ions, and 
a set of six single sided Si strip detectors (SSSD) followed by 2 additional Si PIN diodes (see Fig.~\ref{FigSetup}). 

Each implanted ion 
was identified 
event-by-event through a measurement of its magnetic rigidity, which in connection with 
a time-of-flight measurement provides the mass to charge ratio, and
energy loss in the Si PIN detectors, which provides information about the atomic number. 
The magnetic rigidity was determined by a position measurement at the intermediate dispersive
image of the A1900 fragment separator using a position sensitive plastic scintillator. Time-of-flight was
measured between two plastic scintillators located at the exit of the A1900 and in front of the 
Beta Counting Station (BCS). 

The resulting particle identification spectrum
can be found in reference \cite{Hos05}. The 
implantation detector was a 985 $\mu$m thick DSSD with 40 x 40 
pixels registering time and position of ion implantations and $\beta$ decays.
To increase the $\beta$-decay correlation efficiency, correlations between decays 
and preceding implantations
were established using a 3 x 3 pixel area centered around the implantation location. 
The total implantation rate into all 1600 pixels of the BCS was always less 
than 0.1 ions per second, providing enough time between implantations 
into the same 3 x 3 pixel area for all decays of interest to occur
long before the implantation of the next ion. The SSSD detectors and the downstream 
Si PIN diodes where used to veto beam contaminants that were not stopped in the 
DSSD. 

\begin{figure*}[t]
\includegraphics[width=7cm,bb=300 250 600 500]{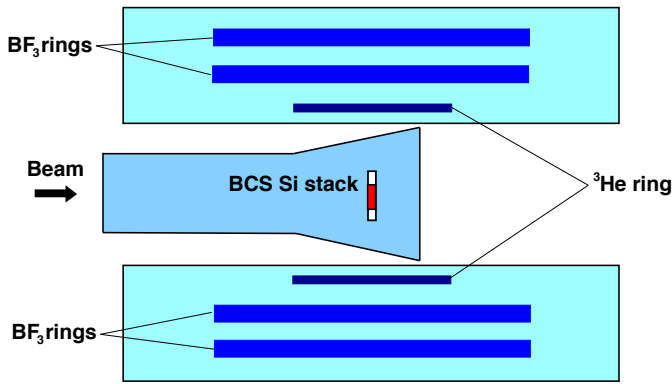}
\caption{\label{FigNERO} Schematic arrangement of the Beta Counting Station (BCS) 
within the NERO neutron detector.}
\end{figure*}

In this paper, we also present results from NERO (Neutron Emission Ratio 
Observer), a neutron detector that surrounded the BCS to provide information 
about branchings for $\beta$-delayed neutron emission \cite{Per09} (see Fig.~\ref{FigNERO} and
also \cite{MEH06,PHA09}). 
NERO is a neutron long counter 
consisting of 60 ionization counters filled with BF$_3$ or 
$^3$He gas imbedded in a block of polyethylene. The counters are arranged
in three concentric rings around the beam axis 
with the inner ring using $^3$He detectors, 
while the two outer rings use BF$_3$ gas counters. 
The center of the detector is a cylindrical cavity with 
22.4~cm diameter containing the beam line and the BCS, 
with the center of the DSSD being located in the center of the NERO detector. 
Neutrons
emitted in the DSSD during $\beta$-decay of a neutron rich isotope
are thermalized in the polyethylene and then detected either by 
a $^{10}$B(n,$\alpha$) reaction in a BF$_3$ gas counter or a 
$^{3}$He(n,p) reaction in a $^3$He gas counter. Neutron detection 
events are registered using a multi hit TDC for each of the 60 
detectors that is started with the detection of the $\beta$-decay event
in the DSSD. 
The typical thermalization timescale for neutrons with initial 
energies of a few MeV (for example for a $^{252}$Cf source) is of the 
order of 70 $\mu$s. To determine whether
a neutron has been emitted after a $\beta$-decay we therefore 
count neutrons for a 200~$\mu$s time period. This 
is sufficient to detect most neutrons (about 96\%) \cite{Per09}. With our low
overall implantation rate of less than 0.1 ions per second this 
is still fast enough to correlate the neutron detection uniquely with 
a beta-decay event and the preceding implantation event.

\section{Results for $\beta$-decay half-lives}

The mixed rare isotope beam used in this experiment contained
$^{73-75}$Co, $^{75-78}$Ni, $^{77-80}$Cu, $^{79-81}$Zn 
as well as smaller amounts of more stable isotopes
$^{81-82}$Ga, $^{76}$Cu, and $^{74}$Ni.  
$\beta$-decay half-lives were determined as described in \cite{Hos05}
using a maximum likelihood analysis method (MLH) \cite{PHA09}. The likelihood function to be maximized
is the product of the probability densities for each implantation event describing
the measured time sequence of 
decay-type events following the implantation within a correlation time $\tau_{\rm corr}$.
The formalism takes into account three decay generations 
and a constant background.  For this analysis
we used $\tau_{\rm corr} = 5$~s. We also take into account
$\beta$-delayed neutron emission for the parent and 
daughter nuclei, which changes the daughter and grand-daughter half-lives. 
The only 
free parameter is the assumed parent half-life. Daughter half-lives and
$P_{\rm n}$ values are fixed and were taken from literature \cite{nubase03} and \cite{Pfeiffer02}
or, when available, from this experiment. 
The advantage of this method is that it avoids the loss of 
information about the exact time, pixel location, and time sequence of the multiple decay events 
following individual implantations of parent nuclei 
that occurs when binning events to form decay curves. 

To determine the detection 
efficiency for $\beta$ particles we performed a traditional
decay curve analysis for the cases with high statistics that 
included $^{75}$Ni, $^{76}$Ni, $^{77}$Cu, $^{78}$Cu, 
$^{79}$Zn, and $^{80}$Zn. The half-lives obtained from the 
fits agree with the ones from the MLH analysis. The 
$\beta$-decay detection efficiencies for the DSSD 
can be determined from the parent decay component 
and the known number of implantation events. The resulting 
efficiencies varied for different isotopes from 40.7\% to 43.0\%, in some cases 
by more than the statistical one sigma errors, but without any obvious trend. For 
the isotopes with less statistics we therefore assumed 
an average $\beta$ detection efficiency of 42$\pm$1 \%
for all three generations. 

The background rate for 
decay-type events was determined for each implantation 
pixel and for each run (typically an hour long)
by counting all decay events outside of a 100~s 
time window following an implantation that occurred in the 
corresponding correlation area. 
Typical background rates  
during the experiment ranged from 
0.008 s$^{-1}$ to 0.015 s$^{-1}$ averaged over the 
entire detector. The largest background rate was found 
near the center of the DSSD, where implantation rates
are highest, with a smooth decrease by about a factor of 10 
towards the detector edges. 
This position and run dependent background rate was
compared with the background rate determined from  
decay curve fits by averaging over the particular
implantation pattern (pixel and run) for the individual 
isotope and agreed reasonably well. 

The resulting $\beta$-decay half-lives together
with the number of implanted isotopes are listed 
in Table \ref{TabResults}. For the convenience of the reader 
we include our previously published half-lives  for 
the Ni isotopes
\cite{Hos05}.   

\begin{table*}
\caption{Number of detected implanted ions $N_{\rm imp}$, half-lives
$T_{1/2}$ from this and previous work, number of detected 
correlated $\beta$-n coincidences  $N_{\beta n}$, 
number of expected background $\beta$ n coincidences from random 
background and daughter decays
$N_{\beta n b+d}$ and $P_{\rm n}$ values from this and previous work
\cite{Pfeiffer02,MGB05a}.
\label{TabResults}}
\begin{tabular}{lrdrlllcc}
Nuclide & $N_{\rm imp}$ & \multicolumn{3}{c}{$T_{1/2}$ (ms)} &  $N_{\beta n}$
  & $N_{\beta n b+d}$ & \multicolumn{2}{c}{$P_{\rm n}$ (\%)} \\
   & & \multicolumn{1}{c}{ {\rm this work}} & \multicolumn{2}{c}{previous} & & & this work & previous \\ \hline

$^{73}$Co &  420 & 41. \pm 6 &  41 & $\pm$ 4			& 4 & 2.1 &  $<$7.9 & $>$9 \\ 		
$^{74}$Co &  331 & 34. ^{+6}_{-9} &  30 & $\pm$ 3		& 16 & 7.9 & 18 $\pm$ 15 & $>26$ \\
$^{75}$Co &    76 & 30. \pm 11 &  		&			& 1 & 1.4 & $<$16 & \\
$^{75}$Ni &  1905 & 344. \pm25 & 600 & $\pm$ 200	& 43 & 16 & 10 $\pm$ 2.8 & \\
$^{76}$Ni &  1441 & 238. \pm18 & 470 & $\pm$ 390	& 43 & 13 & 14 $\pm$ 3.6 & \\
$^{77}$Ni &    159 & 128. ^{+36}_{-32} & 		&		& 13 & 6.2 & 30 $\pm$ 24 & \\
$^{78}$Ni &      13 &  110. ^{+100}_{-60} &	 &		&  3 & 0.8 & & \\
$^{76}$Cu &   277 & 599. \pm 18 & 	641 & $\pm$ 6		& 3 & 2.4 & $<$7.2 & 2.4 $\pm$ 0.5\\			
$^{77}$Cu &  6771 & 466. ^{+21}_{-20} & 469 & $\pm$ 8	& 348 & 35 & 31 $\pm$ 3.8 & 15 $^{+10}_{-5}$\\
$^{78}$Cu &  4653 & 335. \pm 17  & 342 & $\pm$ 11	& 310 & 24 & 44 $\pm$ 5.4 & 15 $^{+10}_{-5}$\\
$^{79}$Cu & 754 & 257. ^{+29}_{-26} & 188 & $\pm$ 25	& 81 & 4.2 & 72 $\pm$ 12 & 55 $\pm$ 17\\
$^{80}$Cu & 16 & 170. ^{+110}_{-50} & 	&			&  0 & 0.1 & & \\
$^{79}$Zn & 2109 & 746. \pm 42 & 995 & $\pm$19		&  19 & 13 & 2.2 $\pm$ 1.4 & 1.3 $\pm$ 0.4\\
$^{80}$Zn & 5043 & 578. \pm 21 & 	545 & $\pm$ 16	& 45 & 40 & $<$1.8 & 1 $\pm$ 0.5\\
$^{81}$Zn & 229 & 474. ^{+93}_{-83} & 290 & $\pm$ 50	& 14 & 4.2 & 30 $\pm$ 13 & 7.5 $\pm$3\\ 
$^{81}$Ga &   75 & 959. ^{+37}_{-29} &	1217& $\pm$ 5	& 1 & 0.2 & $<$21 & 12.1 $\pm$ 0.4\\
$^{82}$Ga & 436 & 610. ^{+83}_{-72} & 599 & $\pm$ 2 	& 21 & 2.5 & 30 $\pm$ 8.0 & 22.3 $\pm$ 0.22\\ 

\end{tabular}
\end{table*}

The errors include statistical and systematic errors. The statistical error is obtained from 
the maximum likelihood analysis. To determine the systematic error we recalculated
half-lives for all possible variations in the input parameters such as efficiency, $P_{\rm n}$ values, and 
half-lives of later decay generations. The envelope of all half-lives with their statistical errors was then used 
as an estimate for the total error. Systematic errors typically represent a small fraction of the total error. 
Contributions of the 
order of 20-30\% of the total error budget are obtained for $^{75-77}$Ni and $^{77-79}$Cu with parent 
$P_{\rm n}$ values and daughter half-lives being the dominant sources.

\section{Results for $P_{\rm n}$ values}

$P_{\rm n}$ values were determined from the number of 
$\beta$-n coincidences $N_{\beta-n}$ detected within a
correlation time $\tau$ after implantation using 

\begin{equation} \label{EqPn}
P_{\rm n} = \frac{N_{\beta n} - N_{\beta n {\rm b}} - N_{\beta n {\rm d}}}
{\epsilon_\beta \epsilon_n \tilde{N}_\beta}
\end{equation}

$\epsilon_\beta$ and $\epsilon_n$ are $\beta$ and neutron 
detection efficiencies respectively. The expected number of 
detected $\beta$-n background events $N_{\beta n {\rm b}}
= r_{\beta n {\rm b}} N_{\rm imp} \tau_{\rm corr}$ 
can be calculated from the background rate 
$r_{\beta n {\rm b}}$, the correlation time $\tau_{\rm corr}$,
and the number of implanted ions $N_{\rm imp}$. The 
number of actual parent $\beta$-decays $\tilde{N}_\beta
= N_{\rm imp}(1-\exp^{-\lambda \tau_{\rm corr}})$
can be determined 
from the known parent $\beta$-decay rate $\lambda$.
$N_{\beta n {\rm d}}$ are neutrons from 
$\beta$-delayed neutron emission of daughter
nuclei, which can be significant in some of the cases
studied here. It can be determined from 
\begin{equation} 
N_{\beta n {\rm d}} =  N_{\rm imp} \epsilon_n \epsilon_\beta \left( (1-P_{\rm n}) n_0 
+ P_{\rm n} n_1 \right)
\end{equation} 
with $n_0$ and $n_1$ given by 
\begin{equation}
n_x = \frac{ P_{nx} \lambda \lambda_x}{\lambda_x - \lambda} 
\left( \frac{1-\exp^{-\lambda \tau_{\rm corr}}}{\lambda} - 
  \frac{1-\exp^{-\lambda_x \tau_{\rm corr}}}{\lambda_x} \right)
\end{equation}
Indices $x=0,1$ indicate the daughters reached without and with neutron emission 
of the parent, respectively. $\lambda$ is the parent decay rate, $\lambda_x$ 
and $P_{nx}$ are daughter decay rates and $P_{\rm n}$ values, respectively. 
Note that $N_{\beta n {\rm d}}$ does depend on the unknown parent 
$P_{\rm n}$ value. Eq.~\ref{EqPn} therefore needs to be rearranged to 
calculate $P_{\rm n}$ (see also \cite{PHA09}). 

The neutron 
detection efficiency $\epsilon_n$  has been measured for a number 
of neutron energies in a separate experiment
at the University of Notre Dame using the $^{13}$C($\alpha$,n),
$^{11}$B($\alpha$,n) and $^{51}$V(p,n) reactions. In the case of 
$^{13}$C($\alpha$,n) and $^{10}$B($\alpha$,n) the neutron 
production was inferred from the well known properties of 
narrow resonances and a measurement of the beam current
using a Faraday cup with electron suppression, 
while in the case of $^{51}$V(p,n) the neutron yield was
determined by offline counting of the induced $^{51}$Cr 
activity. In addition, the detection efficiency was also determined with a calibrated 
$^{252}$Cf source.  Calibration measurements using the same
$^{252}$Cf source were also performed before and after the experiment
reported here. The details of the efficiency calibration are discussed in 
\cite{Per09}, see also \cite{PHA09}. The NERO
efficiency  as a function of  neutron 
energy is constant for low energies and drops for increasing neutron 
energies beyond $\approx$ 1 MeV. The energy 
of each neutron detection event is recorded to monitor correct operation 
of the detector system and setting of the detection thresholds, but does not contain information about 
the initial neutron energy because of the thermalization process. 
Therefore, theoretical assumptions about the maximum neutron energy have to 
be made. From numerous past studies it is well known that $\beta$-delayed 
neutron spectra for medium to heavy nuclei are compressed at energies below
$\approx$800~keV, often below 500~keV (for example \cite{KSO82}, see also more
detailed discussion and references in \cite{PHA09}). The reason is the 
phase space for $\beta$-decay that strongly favors neutron emitting states
just above the neutron threshold, and a tendency for neutron
decay into excited states in the final nucleus. In addition, we used our 
shell model calculations 
\cite{LBH05}  to estimate maximum neutron 
energies for $^{74-76}$Ni, and $^{78-79}$Cu. These calculations predict 
neutron energies in the range of 0.9 - 1.2~MeV when assuming 
the neutron
decay of the daughter states populated by $\beta$ decay
proceeds to the ground state. This assumption obviously leads to an 
overestimation of the neutron energies. We
adopt a conservative upper limit of 1.2~MeV for the neutron energies in 
this experiment, and adopt the corresponding efficiency range
of 33\% to 41\% as systematic error, resulting in 
 $\epsilon_n$=37\%$\pm$4\%.  

The $\beta$-n neutron background rate 
$r_{\beta-n {\rm b}}$ was determined from 
the $\beta$-n coincidence events among the
same $\beta$-type events that were used to 
determine the $\beta$ background. It was
found that across the detector the 
ratio of $\beta$-n to all $\beta$ type events
was constant. We therefore determined this 
ratio for each run, and applied it to the 
position dependent $\beta$-background rate 
used for the half-life analysis to obtain a position 
and run number dependent $\beta$-n 
background rate. The $\beta$-n to $\beta$
event ratio 
was on average about 4\% on the first day of the 
experiment with significant scatter on an hourly 
timescale, 
and increased then to about 7\%
This is about 70 times larger than expected 
from random coincidences with the 
NERO singles neutron background rate 
(which was the same with and without beam)
of 5~s$^{-1}$ and the 
200~$\mu$s $\beta$-n correlation time window. 
A possible explanation is the light ion contamination
in the radioactive beam that deposits 
energies similar to $\beta$ particles in the DSSD
and might emit neutrons when interacting with the 
detector stack. A majority of such events 
were discarded as they pass through the entire
detector stack and were therefore readily vetoed, 
but the remainder might create a 
correlated $\beta$-n background. Such an enhanced 
background was
also observed in a previous NSCL experiment 
with a similar setup \cite{MEH06}. Background from 
random coincidences between actual correlated  parent and daughter 
$\beta$-decays and uncorrelated neutrons
was also considered, but found to be negligible in all cases. 

The resulting $P_{\rm n}$ values are listed together with 
the number of detected $\beta$-n coincidences and
the expected background neutrons in Tab. \ref{TabResults}.
In most cases the statistical error of the number of 
detected neutron events dominates the error entirely. 
For the cases with high statistics, the other major 
contribution to the error is the uncertainty in the 
neutron detection efficiency. It becomes
comparable to the statistical error for $^{78}$Cu and
dominates the uncertainty for $^{77}$Cu. Only in the case
of $^{74}$Co there is 
another significant (more than 10\%) contribution 
to the error. Here the 
uncertainty of the $P_{\rm n}$ value of the daughter, $^{74}$Ni, 
contributes as well. 
Uncertainties in background rate, $\beta$-efficiency,
number of implanted ions, and daughter decay rates
turn out to be negligible in all cases. 

\section{Discussion}
\label{SecDiscussion}

\begin{figure}[h]
\includegraphics*[bb=38 162 285 455,width=6.5cm]{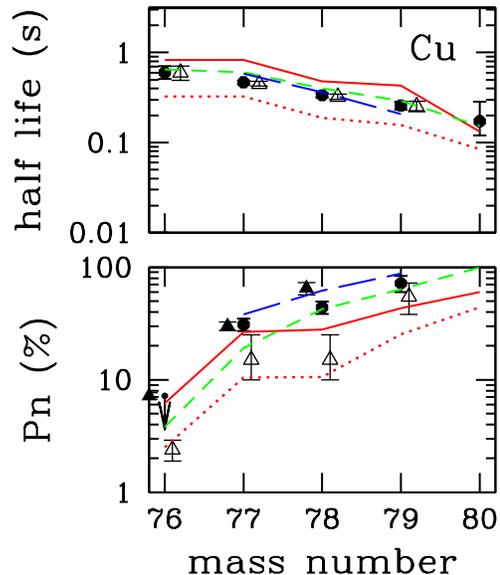}
\caption{\label{FigResultsCu} Half-lives and $P_{\rm n}$ values for the Cu isotopes 
measured in this work 
(solid black circles) compared to previous 
work (open black triangles), and theoretical predictions from QRPA97 \cite{MNK97} (solid red line),
QRPA03 \cite{MPK03} (dotted red line), CQRPA \cite{Bor05} (short dashed green line) and
OXBASH shell model (see text) (long dashed blue line). The experimental results 
from \protect\cite{WIR09} are added as solid black triangles.}
\end{figure}

\begin{figure}[h]
\includegraphics*[bb=38 162 285 455,width=6.5cm]{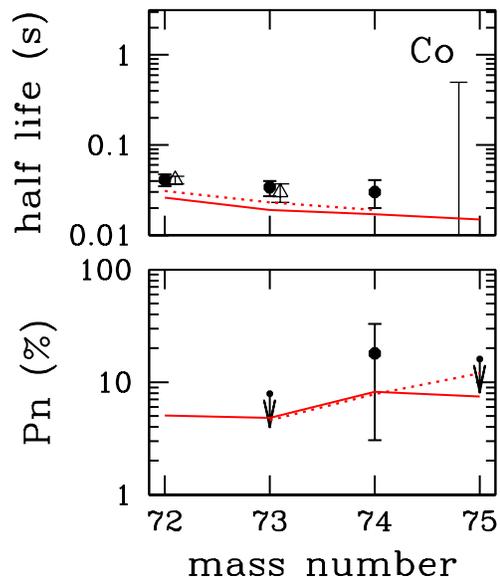}
\caption{\label{FigResultsCo} same as Fig.~\ref{FigResultsCu} for the Co isotopes}
\end{figure}

\begin{figure}[h]
\includegraphics*[bb=38 162 285 455,width=6.5cm]{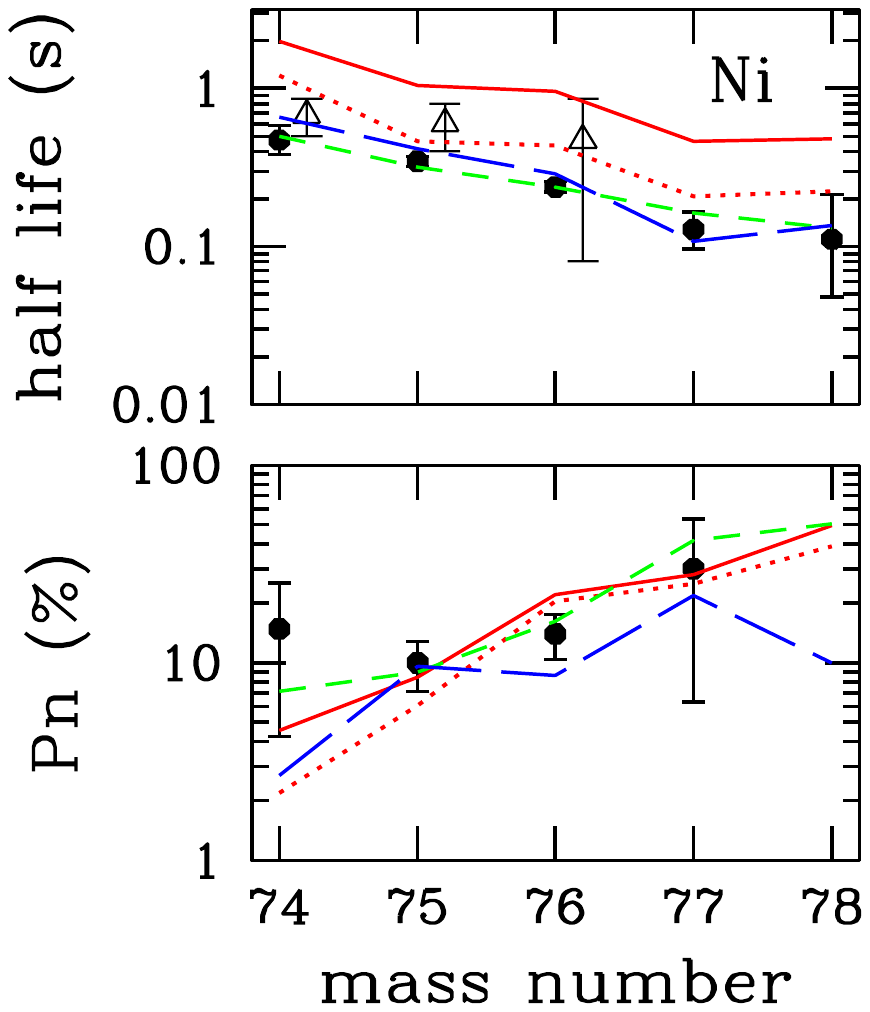}
\caption{\label{FigResultsNi} same as Fig.~\ref{FigResultsCu} for the Ni isotopes}
\end{figure}

\begin{figure}[h]
\includegraphics*[bb=38 162 285 455,width=6.5cm]{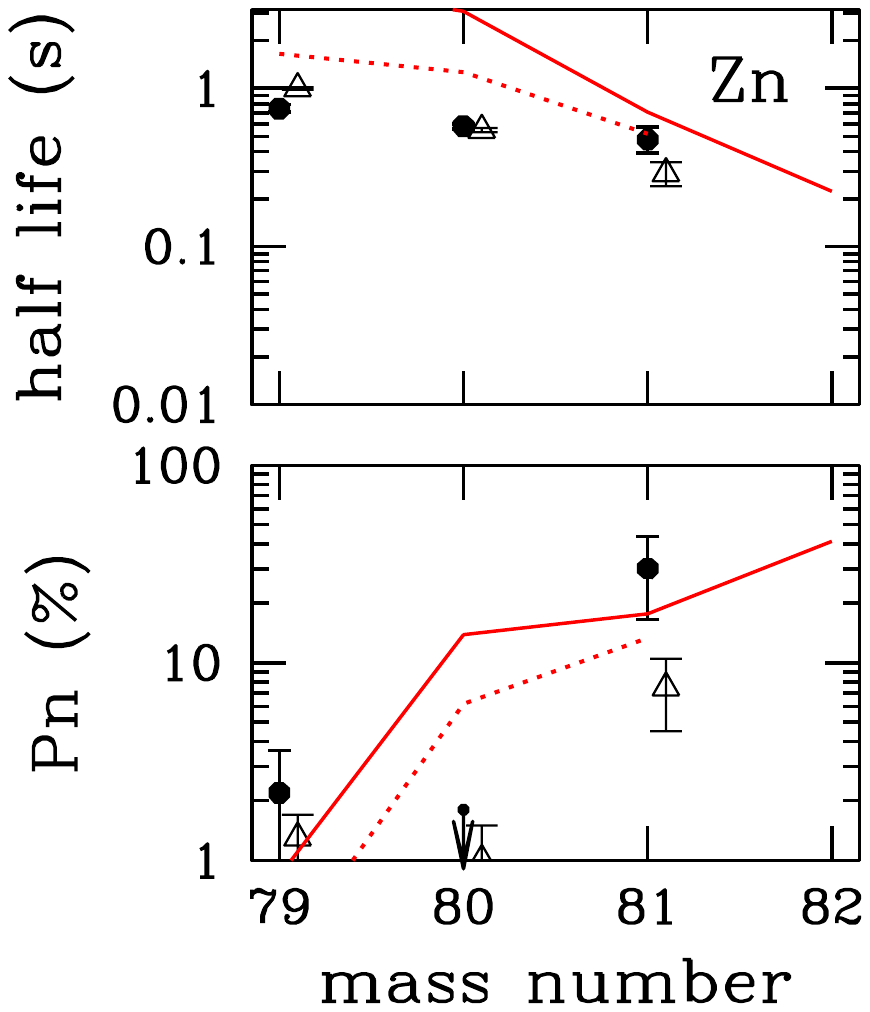}
\caption{\label{FigResultsZn} same as Fig.~\ref{FigResultsCu} for the Zn isotopes}
\end{figure}

\begin{figure}[h]
\includegraphics*[bb=38 162 285 455,width=6.5cm]{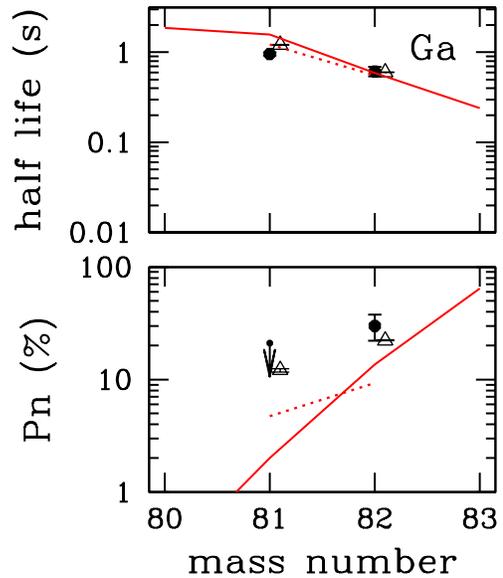}
\caption{\label{FigResultsGa} same as Fig.~\ref{FigResultsCu} for the Ga isotopes}
\end{figure}

Figs.~\ref{FigResultsCu}-\ref{FigResultsGa} show our new half-lives and 
$P_{\rm n}$ values together with previous measurements. 
We include the half-lives of the neutron-rich Ni isotopes that were 
published and discussed earlier \cite{Hos05}.  We now present additional 
half-life data 
for neutron rich Co, Cu, Zn, and Ga isotopes, including the first 
measurement of the half-lives of $^{75}$Co and $^{80}$Cu. In 
most cases there is excellent agreement with previous 
measurements, except for the $^{79}$Zn half-life, for 
which we obtain a significantly shorter value. It has been 
speculated in the past that the existence of an isomer
in $^{79}$Zn cannot be excluded \cite{EFH86}. As the population
of isomers depends on the production mechanism, this could 
lead to differences in measured half-lives between our
technique and previous measurements using the ISOL method
for isotope production. 

In addition to the half-lives, we also obtained
$P_{\rm n}$ values or upper limits for most of the 
isotopes in the beam. For $^{73-75}$Co, $^{74-77}$Ni 
these are the first direct measurements
of $\beta$-delayed neutron emission. In the case of 
$^{77-79}$Cu our measurements are more precise 
and systematically larger than previous work \cite{KGM91}.
As stated by the authors, those experiments at CERN ISOLDE
were difficult due to the chemical non-selectivity of the 
plasma ion source, and have large uncertainties. 
Our $P_{\rm n}$ value for $^{79}$Zn and $^{81}$Ga as well
as the upper limits for 
$^{80}$Zn and $^{81}$Ga are compatible
with previous work. In the case of $^{81}$Zn, our $P_{\rm n}$ value
is larger than the literature 
value \cite{KGM91}. This is in line with new data from ISOLDE
indicating a $P_{\rm n}$ value of larger than 10\% for 
$^{81}$Zn \cite{KBC05}. 

Recently, Mazzocchi et al. \cite{MGB05a} reported 
preliminary lower 
limits for the $P_{\rm n}$ values of $^{73,74}$Co of
9$\pm$4\% and 26$\pm$9\% respectively from 
a $\gamma$-ray spectroscopy experiment. Within
their uncertainties these limits are compatible with our data,
albeit only barely in the case of $^{73}$Co, which would
have to have $P_{\rm n}$ between 5-8\% to be compatible with both, 
our upper limit and their lower limit.  

After completion of this work, new experimental $P_{\rm n}$ values for $^{76-78}$Cu
obtained by $\beta$-delayed $\gamma$-ray spectroscopy have been
reported \cite{WIR09} (see Fig.~\ref{FigResultsCu}). 
The reported value for $^{76}$Cu of  7.0$\pm$0.6~\%
is consistent with our upper limit. For $^{77}$Cu a rather precise value of 
30.0$\pm$2.7~\% was obtained, in excellent agreement with our result. 
For $^{78}$Cu the reported result of 65$\pm$8~\% is significantly larger 
than our measurement. This might reflect the difficulty of obtaining a
reliable $P_{\rm n}$ value via $\gamma$-ray spectroscopy in this particular case
as indicated by the authors \cite{WIR09}. 

In Figs.~\ref{FigResultsCu} - \ref{FigResultsGa}  our experimental results are compared
with various theoretical calculations. 
The  shell model calculations  have been performed using the JJ4B effective 
interaction \cite{LB08} for the $p_{3/2}f_{5/2}p_{1/2}g_{9/2}$ model space on top of a $^{56}$Ni 
inert core. JJ4B has been developed starting from a realistic G-matrix interaction based on
the Bonn-C NN potential together with core-polarization corrections \cite{LBH04}.
It has been fitted to reproduce new experimental data separately for $^{57-78}$Ni isotopes
 (purely neutron interaction) and for $N=50$ ($^{79}$Cu-$^{100}$Sn) isotones (purely proton interaction). This interaction has been extended further to incorporate the proton-neutron degree of freedom. The new 
 version of the effective interaction JJ4B \cite{LB08} takes into account information about 450 
states in 73 nuclei including recent experimental data for Cu, Zn, Ga, Ge, As, Se, Br and Kr in the vicinity 
of $^{78}$Ni. 

The number of chosen orbitals in these calculations is sufficient to 
achieve a satisfactory description of binding 
energies, neutron separation energies and known excitation spectra for neutron-rich
nuclei in the considered region. However, we find that to reproduce 
the half-lives and P$_n$-values of neutron-rich Ni, Zn and Cu isotopes the 
GT operator has to be renormalized by a factor of 0.37, instead of the
expected factor of 0.75 for this mass region. Such a strong renormalization 
of the GT operator indicates that shell model configurations responsible for 
considerable amounts of GT strength are not accounted for. Such configurations can be 
attributed to the excluded f$_{7/2}$ orbital, which is connected to its spin-orbit 
partner, f$_{5/2}$, by a very strong GT matrix element.  Indeed, a test calculation in an
enlarged model space that includes the proton f$_{7/2}$ orbital and uses a $^{48}$Ca core 
has been performed with a combined effective interaction \cite{LBS06} and confirms
 that the exclusion of the f$_{7/2}$ orbital in the model 
space represents a strong limitation. This demonstrates the importance of 
$\beta$-decay data, including $P_{\rm n}$ values, 
in testing shell model calculations far from stability. 

A global nuclear structure
model is needed for astrophysical applications that is not limited to nuclei within a 
specific model space or near shell closures. We therefore
compare our data to theoretical results from the global, universal 
quasi-particle random-phase approximation (QRPA) model
the details of which can be found in 
Refs.~\cite{krumlinde84:a,moller90:a,MNK97,MPK03}. This model 
considers Gamow-Teller $\beta$-decay transitions from a parent 
nucleus to the accessible states in the daughter. 
The starting point is to
calculate nuclear wave functions for the ground-state shape of the nucleus in a
folded-Yukawa single-particle model. Ground state shapes are 
taken from the finite range droplet model (FRDM) \cite{moller95:b}.
The decay rates are obtained as matrix elements of the 
Gamow-Teller operator between parent and daughter states in a QRPA 
with pairing and Gamow-Teller
residual interactions. A global table of calculated $\beta$-decay half-lives
and $\beta$-decay delayed neutron emission rates was published in
1997 \cite{MNK97}. For that calculation
$\beta$-decay Q-values had been obtained from
the 1989 Atomic Mass Evaluation \cite{audi89} when available, 
otherwise from the FRDM \cite{moller95:b}
In our comparisons to experimental data we denote this theoretical
data set QRPA97.

Subsequently several 
enhancements have been made to this model resulting in a new dataset QRPA03 \cite{MPK03}:
An empirical spreading was applied to the Gamow-Teller
strength function, and for nuclei near magic numbers an exact
spherical shape was assumed instead 
of the weakly deformed shapes
obtained for these nuclei in the FRDM \cite{moller95:b}. In addition, a first-forbidden
strength distribution as predicted by the gross theory 
\cite{takahashi72:a,takahashi73:a} was added. Compared  to the
allowed Gamow-Teller strength, which over a given energy range is represented by
relatively few strong peaks, the first forbidden strength with its numerous
small densely spaced peaks to a good
approximation constitutes a ``smooth background''. It is therefore a reasonable
approach to calculate the GT transitions in a microscopic QRPA approach
and the {\it ff} transitions in a macroscopic statistical model, in analogy with
the macroscopic-microscopic method used for mass models. 
The experimental masses used to determine the $Q_{\beta}$ value
were from \cite{AME95} when available.

We can now compare both, QRPA97 and QRPA03, with our new 
experimental data in the Co-Zn region. 
Overall QRPA03 agrees better with our set of measurements, with the exception of the 
Cu isotopes, where compared to QRPA97 discrepancies increased, especially for the 
$\beta$-delayed neutron-emission probabilities
(Figs.~\ref{FigResultsCu}). 

\begin{figure*}[t]
\includegraphics[width=10cm]{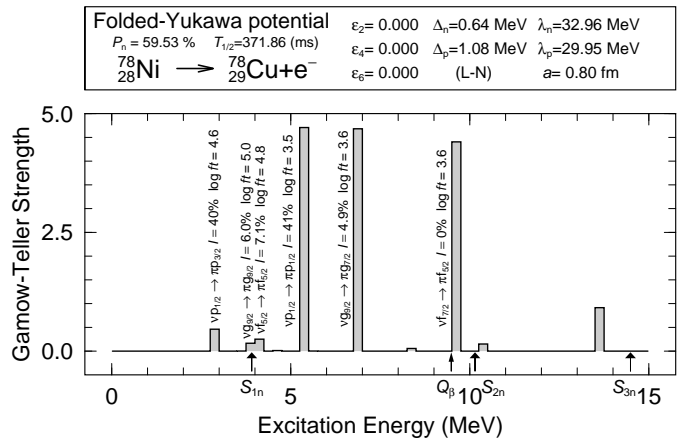}
\caption{\label{str}Calculated Gamow-Teller $\beta$-strength
for $^{78}$Ni. We indicate by wide arrows the 1n, 2n, and 3n
neutron separation energies in the daughter $^{78}$Cu following
the decay of $^{78}$Ni. Only states below $Q_{\beta}$
are accessed in the decay. The figure is further discussed
in the text.}
\end{figure*}

There are several possibilities
for the origin of the discrepancies: The
calculated half-life depends critically on the $Q_{\beta}$. For example,
for $^{78}$Ni, a change of  $Q_{\beta}$ of 1~MeV would change the 
half-life by roughly a factor of 2. As the half-life depends on a few 
low lying transitions, a 1~MeV change in the main low lying 
$\beta$-strength just below 3~MeV (see Fig.~\ref{str}) would have 
a similar effect. $P_{\rm n}$ values can be even more sensitive to 
the exact location of transitions in the strength distribution, in particular 
in a case like $^{78}$Ni, where there is significant strength right around 
the one neutron separation energy. Therefore, calculated
$P_{\rm n}$ values are extremely sensitive to small variations
in the calculated strength distribution near $S_{\rm 1n}$, and no model
can predict accurately the energy levels or strength at these
relatively high energies.

In addition, the strength distribution depends on 
deformation and on whether one includes first forbidden transitions. This 
is illustrated in Table \ref{tab2}.  Clearly, when the GT-only transitions are fast
(the first three rows) then adding ff transitions has a small
effect. Fast GT transitions correspond to low-lying GT strength. Therefore
adding a small ff component at these energies has little effect.
In the case of long half-lives (last four rows) the effect
of including ff transitions is more substantial.  One could perhaps argue that a standard single-particle level
diagram for this region of nuclei reveals no obvious candidates
for such first forbidden decays.  
Fig. \ref{str}
indicates the main single-particle component of the strongest transitions.
However, we should recall that the residual pairing and Gamow-Teller
interactions considerably change both the energy and wave-function structure
from the simple single particle picture. Therefore a single-particle level
diagram can only provide rough guidance to $\beta$-decay properties.
More realistic nuclear interactions
might well yield a level structure that contains first forbidden decays
to low-lying states.We also note that the effect of
even fairly weak deformation can be significant.

It is therefore noteworthy that the global
model still reproduces measured $P_{\rm n}$ values to within a factor
3. The good global agreement between calculated and measured
$P_{\rm n}$ and $T_{1/2}$ values is obtained although {\it no model parameter}
was varied to adjust the model to these experimental quantities.

\begin{table*}
\caption[tabbb]{\label{tab2}Calculated $\beta$-decay half-lives and delayed-neutron-emission
probabilities for four assumptions: 
(1) GT transitions and calculated ground-state shape, 
(2) GT transitions and spherical ground-state shape, 
(3) GT+ff transitions and calculated ground-state shape, and 
(4) GT+ff transitions and spherical ground-state shape.}
\begin{tabular}{rrrrrrrrrrrrrr}
\hline
\hline  \\[-2.1ex]
& & &
   \multicolumn{5}{c}{$T_{1/2}$ (ms)}  & &
   \multicolumn{5}{c}{$P_{\rm n}$ (\%)}    \\
\cline{4-8} \cline{10-14} \\[-2.1ex]
 & & & 
\multicolumn{2}{c}{GT}    & &
\multicolumn{2}{c}{GT+ff} & &
\multicolumn{2}{c}{GT}    &&
\multicolumn{2}{c}{GT+ff} \\
\cline{4-5} \cline{7-8} \cline{10-11} \cline{13-14} \\[-2.1ex]
   \multicolumn{1}{c}{$Z$}              & 
   \multicolumn{1}{c}{$A$}              & 
   \multicolumn{1}{c}{$\epsilon_2$}     &
Def. & Sph. & &
Def. & Sph. & &
Def. & Sph. & &
Def. & Sph. \\
\hline \\[-2.1ex]
   27 &  73 &  0.092  &     27 &    33 & &   25 &    30 & &    5.34&   2.82& &  6.94&   4.59\\
   27 &  74 &  0.117  &     18 &    26 & &   16 &    22 & &    9.30&   4.75& & 10.19&   7.76\\
   27 &  75 &  0.092  &     15 &    20 & &   13 &    18 & &    7.17&   5.68& & 10.41&  11.98\\
   28 &  75 &  0.058  &   1053 &   818 & &  539 &   460 & &    9.93&  13.27& &  5.86&   6.06\\
   28 &  76 &  0.050  &   1064 &   674 & &  585 &   433 & &   25.94&  36.18& & 16.97&  20.49\\
   28 &  77 &  0.050  &    428 &   370 & &  226 &   207 & &   33.58&  35.16& & 22.80&  25.17\\
   28 &  78 &  0.025  &    371 &   371 & &  228 &   224 & &   38.26&  59.53& & 28.69&  39.06\\
   29 &  76 &  0.117  &    734 &   463 & &  435 &   318 & &    8.06&   1.28& &  4.74&   2.46\\
   29 &  77 &  0.083  &    827 &   417 & &  505 &   318 & &   32.81&  10.40& & 19.30&  10.38\\
   29 &  78 &  0.075  &    386 &   289 & &  224 &   187 & &   34.82&  11.16& & 19.79&  10.60\\
   29 &  79 &  0.050  &    391 &   224 & &  223 &   155 & &   46.32&  18.97& & 38.07&  25.24\\
   29 &  80 &  0.075  &    103 &   160 & &   64 &    84 & &   58.46& 100.00& & 51.53&  56.62\\
   30 &  79 &  0.067  &   4991 &  3591 & & 1869 &  1647 & &    1.29&   1.76& &  0.29&   0.34\\
   30 &  80 &  0.042  &   3796 &  2505 & & 1580 &  1259 & &   20.91&   4.04& &  5.72&   6.19\\
   30 &  81 &  0.075  &    707 &  3160 & &  325 &   517 & &   19.53&  43.42& & 11.34&  13.38\\
   31 &  82 &  0.083  &    539 &  2062 & &  304 &   553 & &   14.43&  34.76& &  9.03&   9.32\\
\hline
\hline
\end{tabular}
\end{table*}
 
Nevertheless the global QRPA calculations
do show a significant underprediction of the high $P_{\rm n}$
values for the Cu isotopes found in this experiment and in \cite{WIR09} 
pointing to some issue in this model for this particular region. 
As discussed
above, there are a number of possible reasons that could lead to such 
a deviation. One such possibility is uncertainty in the nuclear masses. 
Since QRPA03, the masses of the Zn daughter isotopes out to $^{81}$Zn 
have been determined with high precision Penning Trap measurements \cite{BAB08,Hak08}.
As we do not mix theoretical and experimental masses to determine 
Q-values this does not affect the $\beta$-decay Q-values, but it allows
for a precise determination of the neutron separation energies of the Zn isotopes, which 
are needed to determine the Cu $P_{\rm n}$ values. A recalculation with 
these new masses lead only to 10-30\% changes compared to QRPA03, 
by far too small to explain the observed discrepancy. Furthermore, in the case of 
$^{76}$Cu both, $\beta$-decay Q-value and daughter neutron separation 
energies are known experimentally with keV precision in the QRPA03 calculation.
Clearly, the underprediction of $P_n$ (with the new value of 7.3 $\pm$ 0.6 \% \cite{WIR09})
and half-life is already present in $^{76}$Cu. Therefore, masses are unlikely to 
be the explanation for this problem. 
The extensive $\gamma$-ray data from \cite{WIR09} for $^{77}$Cu,
once published, might help to test calculated strength functions in more detail
to determine the cause of this problem. 

Our recalculation of the QRPA predictions with updated Zn masses had
significant impact on the predicted Zn half-lives, where the $\beta$-decay 
Q-values are now known experimentally. For example, for $^{79}$Zn the 
$\beta$-decay Q-value increases from 8.68~MeV as predicted by the FRDM to 
9.08$\pm$0.1~MeV, decreasing the predicted half-life from 1.6~s to 1.0~s. 
Similarly, the $^{80,81}$Zn half-lives decrease from 
1.3~s, and 0.52~s to 1.0~s and 0.35~s, respectively. Overall this reduces the 
discrepancy between experiment and theoretical prediction significantly. 

Finally, we compare our new data with the continuum QRPA (CQRPA)
calculations from Borzov \cite{Bor05}. His treatment is limited
to spherical nuclei and is not global, but does  include a pn interaction 
in the particle-particle channel and a microscopic calculation of the 
first forbidden strength.  Table \ref{tab2} shows that the assumption
of spherical shapes may be inappropriate as even 
the modest deformations predicted for his region are expected to 
have a substantial
effect on half-lives and $P_{\rm n}$ values. 
The predictions of half-lives and $P_{\rm n}$ values by Borzov
show overall very good agreement with our experimental results. Borzov 
finds, that in his model first forbidden transitions play only a minor role 
around $^{78}$Ni as long as $N \leq 50$.
However, the CQRPA model does significantly underpredict the $P_{\rm n}$ values of $^{76}$Cu 
and $^{77}$Cu, while there is excellent agreement for $^{78}$Cu and $^{79}$Cu. 

Motivated by their new experimental $P_{\rm n}$ values for $^{77-78}$Cu Winger et al. \cite{WIR09} 
argue that  this underprediction
is an indication for the inversion of the $\pi 2p_{3/2}$ and $\pi 1f_{5/2}$ single particle 
orbitals. CQRPA model calculations where these orbitals are inverted
indeed lead to larger $P_{\rm n}$ values, though no half-life data are presented that 
would allow to verify consistency with experimental half-lives. However, 
our $P_{\rm n}$ value for $^{78}$Cu is significantly lower than the 
experimental value reported by 
\cite{WIR09}. While the old CQRPA results were in agreement with our 
measurement,  the new CQRPA value of 53\% is somewhat high. In 
addition, the old
CQRPA results are also in excellent agreement with our new $P_{\rm n}$ value 
for $^{79}$Cu.  Nevertheless, there is other experimental evidence for such a level inversion 
to occur for Cu isotopes at $^{75}$Cu and beyond \cite{Flan09} thereby justifying the 
modifications of the CQRPA model. 

\section{r-process calculations}

With our new data and recent precision mass measurements around 
$^{80}$Zn \cite{BAB08,Hak08} the nuclear physics needed to model 
the r-process around $A=80$ is now to a large extent experimentally 
determined. We can therefore test r-process models in this particular
mass region against observations with greatly reduced nuclear physics uncertainties. 

 It is quite challenging 
to understand the origin of the elements in this mass region, 
as not only all major neutron capture processes, the weak and strong
s-process and the r-process, can contribute, but charged particle 
reaction sequences can reach this mass region as well. Indeed,
one class of r-process models, the neutrino driven wind scenario
in core collapse supernovae, predict that nuclei in this region 
are produced by a combination of charged particle and 
neutron induced reactions. Nevertheless, we can ask,
 whether r-process models characterized by a neutron 
capture and $\beta$-decay reaction sequence in the $A=80$ mass region 
are now able to reproduce the observed solar r-process abundances
in this region, or not.  

To address this question we use a classical r-process model 
that simulates a series of neutron exposures of Fe seed nuclei with neutron density 
$n_n$, temperature $T$, duration $\tau$, and weight $\omega$. It has been 
shown that such a model can reproduce the observed solar 
abundance pattern reasonably well employing power law
relationships $\omega (n_n)=a_1 n_n^{a_2}$ and 
$\tau (n_n)=a_3 n_n^{a_4}$ leaving only 
3 free parameters plus an overall normalization \cite{Cow99}. The temperature
$T=1.35$~GK  is kept constant and is the same for all components representing a 
typical r-process freezeout temperature. The model we use
adopts the waiting point approximation and assumes 
a sudden freezeout with decay back to stability once 
all exposures have been applied. The fact that the 
site of the r-process is still unknown and that a wide range
of scenarios have been proposed, motivates the use of this 
simple site-independent model. The classical model can simulate well the final local 
neutron capture flow in an r-process scenario at 
freezeout, which tends to dominate the features of the 
final abundance pattern. Of course further modifications of the abundance
pattern can occur during freezeout, but this effect is highly model dependent.
It therefore makes sense to explore the agreement of a simple r-process model 
with observations. Once the nuclear physics is fixed, major disagreements with 
observations might indicate an entirely different r-process mechanism for
the mass region in question,
while smaller deviations might reveal additional, site specific effects, such as an 
extended freezeout.   

\begin{figure}[h]
\includegraphics*[bb=38 164 370 415,width=8.5cm]{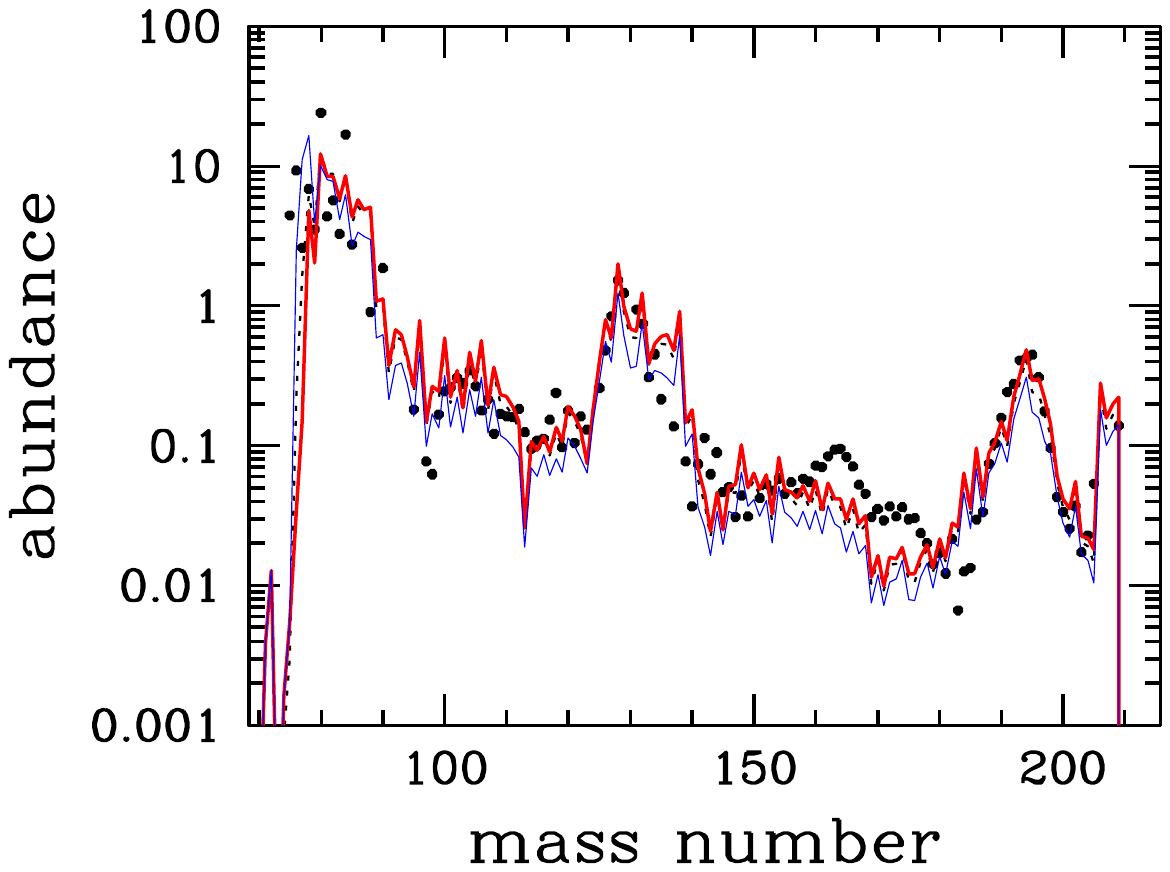}
\includegraphics*[bb=45 164 380 415,width=8.5cm]{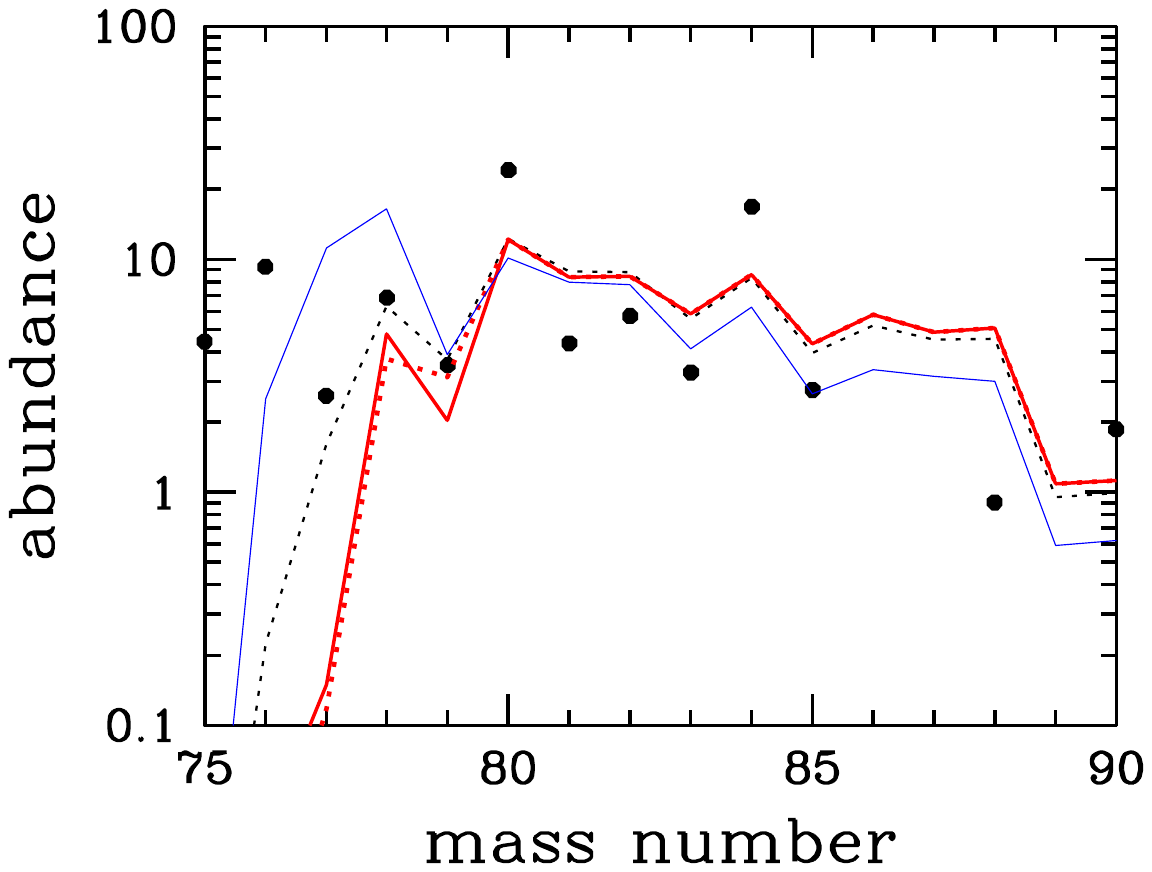}
\caption{\label{FigRproc} Abundances calculated with a classical r-process model
using previously known decay data (dotted black line) and the new data from 
this work (thick solid red line). The thin solid blue line indicates a calculation with 
the large $^{78}$Ni half-life predicted by QRPA97. The thick dotted red line shows 
results using the new half-lives from our work, but the old $P_{\rm n}$ values. Solar
r-process abundances are shown as black data points. }
\end{figure}

Fig.~\ref{FigRproc} shows results from an r-process calculation, where the four 
model parameters have 
been fitted to reproduce the solar r-process 
residuals from Arlandini et al. \cite{AKW99}. These r-process 
residuals have been obtained  by subtracting 
from the observed solar abundances a main s-process component 
calculated with a classical s-process model and a weak 
s-process component. 
In the $A=80$ mass region this solar r-process abundance distribution is very similar
to the one obtained with a set of realistic stellar s-process models
and updated solar abundances \cite{TGA04}. Our baseline model 
calculation with model parameters $a_1 = 4224.74$, $a_2=-0.082$,
$a_3=788.86$ and $a_4=0.0089$
uses the nuclear masses from the 2003 Atomic
Mass Evaluation \cite{AME03}, 
including their extrapolations, updated with the new
masses in the Zn region \cite{BAB08,Hak08}. Masses of 
more exotic nuclei are taken from the ETFSI-Q mass model,
which has been shown to be well suited for r-process 
calculations. We also performed calculations using the FRDM mass model 
that lead to very similar results in the $A=80$ mass region and
to the same conclusions. 
Experimental $\beta$-decay data, including $P_{\rm n}$ values,
are taken from NNDC \cite{Wallet07} when available, otherwise
our theoretical QRPA03 data are used. For comparison
we then run the same calculation with our updated 
$\beta$-decay properties from this study. We also 
show a calculation using the large $^{78}$Ni half-life
predicted in our earlier QRPA97 model. 

The global effects of implementing our new experimental 
data are rather modest. However, the half-life of $^{78}$Ni 
clearly plays an important role, and the large $^{78}$Ni 
half-life of QRPA97 leads to noticable changes across all 
masses, including the $A=195$ peak region. Our 
measurement of the half-life of $^{78}$Ni now excludes
such a long half-life with certainty. 

In the $A=80$ region changes are more significant. 
Our new data lead to a more pronounced odd-even effect 
for $A=78,79,80$ that agrees better with observations. 
As Fig.~\ref{FigRproc} shows, our shorter $^{78}$Ni half-life
actually leads to a weaker odd-even effect between 
$A=78$ and 79, but the larger $P_{\rm n}$ values for 
the Cu isotopes, especially $^{79}$Cu, more than compensate
and are therefore important to obtain a better fit. 
The need for large $P_n$ values, especially for $^{79}$Cu in order
to reproduce the pronounced odd-even effect in the 
r-process abundances has already been pointed out
in earlier studies based on theoretical predictions of half-lives and $P_n$ values
\cite{KHH90}. 

Overall, with the masses and decay data for $A \le 81$ now available
the $A=80$ region can be reproduced 
quite well. The agreement is very good
for $A=78,79,80$ where some experimental data are now available. 
On the other hand, 
some problems become apparent in the 
$A=81-90$ mass range. Because of the lack of experimental 
constraints on the nuclear physics in this region, a nuclear
physics explanation cannot be excluded yet. In this 
region, most of the relevant masses, and most of the 
relevant $\beta$-decay half-lives beyond $A=84$ are 
unknown. Clearly more experimental work 
needs to be done to extend the mass region of reliable
nuclear physics over the entire $A=80$ abundance peak
area.

Even in the $A=78-80$ area there are still some remaining nuclear
physics uncertainties. The most important nuclear physics
data at the $N=50$ shell closure are the half-lives of the major
$N=50$ waiting points, $^{78}$Ni, $^{79}$Cu, and $^{80}$Zn 
and the nuclear masses for the isotopic chains,
where the neutron capture flow might begin
to cross the $N=50$ shell closure.
Owing to the rather large drop in neutron separation energy 
across the $N=50$ shell gap it is clear that regardless of the mass
 model adopted, for the Ni and Cu isotopes the neutron densities 
 required for the reaction flow to cross $N=50$  ($n_{\rm n} > 10^{24,26}$ for the Cu, Ni 
isotopic chains, respectively) are much higher than the
neutron densities that produce the bulk of the $A=80$ isotopes. Therefore, 
for the calculation of abundances around $A=80$, precision masses to 
characterize the breakout beyond $N=50$ in detail are mainly needed for 
Zn and Ga. For these isotopic chains masses are needed 
across  $N=50$ out to $N=52$ because
of the odd-even effect of the neutron separation energies. 
In addition, $P_{\rm n}$ values are needed for all $A=78-80$ waiting points
and their decay daughters. 
With our data and previous work, all of
these quantities are now experimentally known, except for
the mass of $^{82}$Zn, which might introduce some uncertainty 
concerning the $^{80}$Zn waiting point. This has 
been discussed in detail in \cite{BAB08}. In addition, the 
measured $^{78}$Ni half-life still has a large error bar.  
While the rather long half-life predicted by QRPA97 is now 
excluded, the half-life change that leads to the changed
$A=78$ abundance indicated in Fig.~\ref{FigRproc} is 
actually of the order of the experimental uncertainty. Therefore,
a more precise $^{78}$Ni half-life would be desirable.
The $P_{\rm n}$ value of $^{78}$Ni might have some influence on the 
result as well, and should be determined experimentally. 

Finally, we also explore the role of the nuclear physics in 
the $^{78}$Ni region in a more site specific r-process model, 
the so called high entropy wind (HEW) scenario
\cite{WWH92,TWJ94,MeB97,FRR99, WKM01,TBM01,BLD06,FKM09}. The model is 
 inspired by the conditions expected near the proto-neutron star forming 
in a core collapse supernova shortly after the explosion. The high neutrino luminosity
is thought to drive outflows of strongly heated, low density matter (hence high entropy)
that at late times
become neutron rich. While this is one of the most promising 
candidates for an r-process scenario, realistic models based
on conditions obtained in current supernova models do not lead
to a full r-process. We therefore use the simplified parametrized model
of \cite{FRR99,FKM09} that follows a set of one dimensional adiabatic expansions
(components), 
each characterized by an entropy per baryon $S$, an initial 
electron abundance $Y_e$, and an expansion 
velocity $v$. Nuclear reactions are followed
with a full reaction network including $\beta$-decay properties and all neutron, 
proton, and $\gamma$ induced reactions, but neglecting neutrino interactions
and fission. One possible choice of  parameters that lead to a 
successful r-process is to keep the same realistic 
values of $Y_e = 0.45$ and $v=7500$ km/s for all components, but  choose 
an equidistant set of entropies that ranges up to about $S/k\sim 250$, larger than predicted by supernova models. 
When assuming that equal amounts of material are processed by each component, 
such a model has been shown to reasonably reproduce the solar r-process 
residuals \cite{FRR99,FKM09}. We use the same $\beta$-decay rates
and $P_n$ values employed in our classical model calculations. Neutron and 
charged particle induced reaction rates are taken from NON-SMOKER
statistical model predictions \cite{NON-SMOKER} using the FRDM mass model. $\gamma$
induced reactions are calculated from their inverse capture reactions via
detailed balance. 

\begin{figure}[h]
\includegraphics*[bb=30 160 370 415,width=8.5cm]{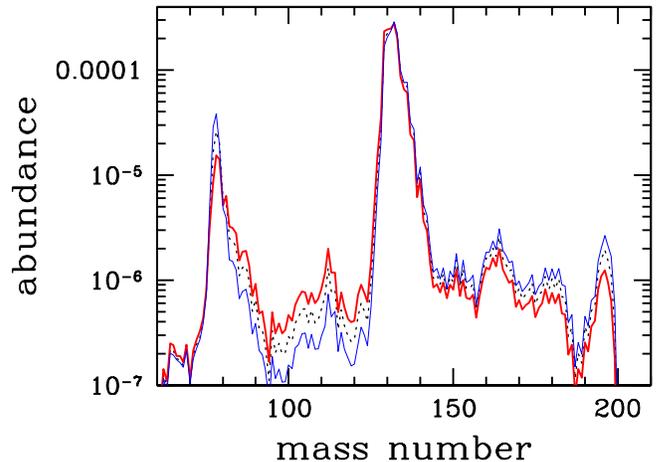}
\caption{\label{FigHEWs200} Abundances calculated with the HEW r-process model
for a single entropy component with $S/k=$200. Results based on the data from this work ($^{78}$Ni half-life of 110~ms) (thick solid red line)  are compared with results based on previously available data  ($^{78}$Ni half-life of 224~ms)  (thin solid blue line) and
previously available data with the long $^{78}$Ni half-life of 477~ms from \protect\cite{MNK97} (black dotted line). }
\end{figure}

\begin{figure}[h]
\includegraphics*[bb=45 160 370 415,width=8.5cm]{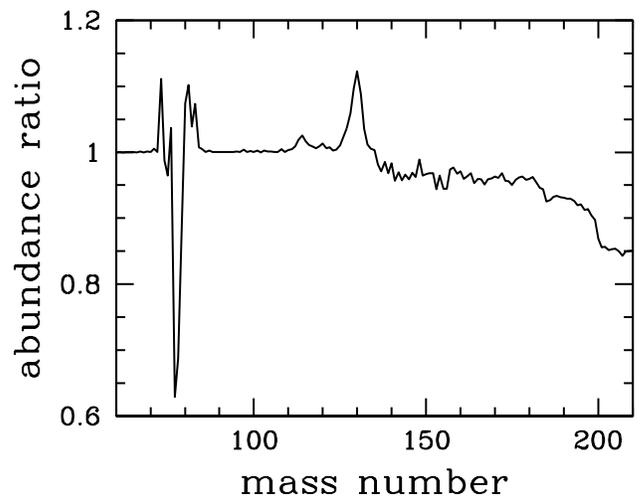}
\caption{\label{FigHEWratio} Ratio of final HEW abundance distributions summed over all 
entropy components calculated with our new
 data including a shorter $^{78}$Ni half-life of 110~ms to what one obtains using previously available data and the long $^{78}$Ni half-life of 477~ms from \protect\cite{MNK97}. }
\end{figure}

In HEW models, r-process seed nuclei are produced by combined charged 
particle and neutron induced  processes in the $A\sim 90$ region
close to stability. As this is already beyond the $^{78}$Ni 
region, one might expect that $\beta$-decay properties near $^{78}$Ni do not 
play a role. Our calculations show, however,
that at high entropies around $S/k=200$, where neutron to seed ratios
become high enough to produce the heavier r-process elements, 
$^{78}$Ni becomes part of the r-process path. Fig.~\ref{FigHEWs200} shows
the abundance distribution for $S/k=200$ with the previous 
nuclear database, and with our new experimental results. It turns out 
that the change is entirely due to our new, shorter $^{78}$Ni half-life (110~ms
instead of 224~ms). 
For comparison we also show a calculation with the older, longer
$^{78}$Ni half-life (477~ms) predicted by \cite{MNK97}. Clearly a long $^{78}$Ni 
half-life reduces the $A=78$ production, but increases the production 
of $A=100-120$ nuclei at $S/k \sim 200$. Interestingly, the production of very heavy 
r-process nuclei is slightly suppressed by a shorter $^{78}$Ni half-life, contrary 
to what one would expect naively and opposite from the behavior in the 
classical model. Nevertheless, the impact of the $^{78}$Ni half-life on the 
final abundances, once all entropy components have been added up, is 
rather small. 
Fig.~\ref{FigHEWratio} shows the ratio of our new abundances
to what one obtains with the long $^{78}$Ni half-life from \cite{MNK97}. 
Besides the significant change at $A=78$, there is a 10\% increase at $A=130$
(where $S/k=$200 makes its largest contribution) and a ~10\% suppression of 
very heavy nuclei. This modest sensitivity of the final abundances
reflects the rather narrow entropy range between $S/k \sim 190-210$ that is broadly 
influenced by  $^{78}$Ni.  Effects for 
higher entropies are still significant but only below $A\sim 130$ where those 
high entropy components do not contribute much. 
However, it is likely that critical waiting points such as 
$^{78}$Ni in the case of $S/k\sim 200$
 do exist also for the other entropy components. It will be important 
to identify and measure these waiting points to obtain more reliable HEW 
r-process calculations. In addition to the already
known $N=50$ isotopes $^{80}$Zn and $^{79}$Cu, our first measurement of the $^{78}$Ni
half-life is an important further step towards this goal.

\begin{figure}[h]
\includegraphics*[bb=30 162 370 415,width=8.5cm]{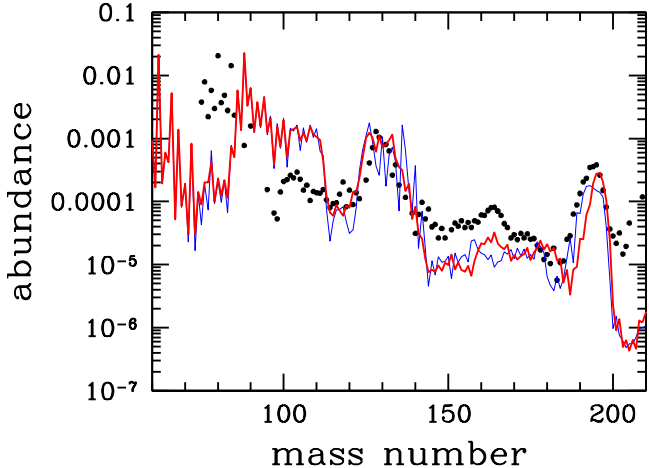}
\caption{\label{FigHEWnopn} HEW abundance distributions summed over all entropy components
using our new data (thick solid red line) and setting all $P_{\rm n}$ values to zero
(thin solid blue line) compared with the solar r-process residuals (black filled circles). }
\end{figure}

Using our new $P_n$ values instead of the old ones does not 
lead to significant changes in the
calculated abundances for our choice of HEW model parameters. In principle one can expect 
a reduced impact of $\beta$-delayed neutron emission compared to the 
classical model, as neutrons are present at later times potentially 
reversing the effects via neutron capture. Once the neutrons are exhausted, 
the r-process path tends to be already closer to stability where $P_n$ values
are smaller. To explore the impact 
of $P_n$  values in HEW calculations, we
run a simulation without any $\beta$-delayed neutron emission for comparison. 
This is similar to what has been done in \cite{KBT93} for the classical 
r-process model. The result is shown in Fig.~\ref{FigHEWnopn} and demonstrates the importance 
of $\beta$-delayed neutron emission in HEW models. For  $A<110$ the impact is less pronounced, though for some mass chains significant changes of up to a factor of 2 do occur. On the other hand,
$P_n$ values play a critical role in shaping the $A=130$ and 
$A=195$ abundance peaks as well as the rare earth peaks. 

Overall the impact of our new data on the synthesis of $A \sim 80$ r-process nuclei 
in our HEW model is rather modest, in contrast to what we found in the classical model. 
However, as has been discussed extensively in \cite{FRR99}, and
as can be seen in Fig.~\ref{FigHEWnopn}, our particular choice of HEW model parameters, 
especially our initial $Y_e$, clearly does not 
reproduce the solar r-process residuals below $A \sim 110$. Once this problem 
has been solved, the question of the relevance of the nuclear physics in the $A=80$
region in HEW models should be revisited. 

\section{Summary}
We have provided the first experimental $P_{\rm n}$ values for extremely 
neutron rich Ni isotopes out to $^{77}$Ni and a first
half-life for $^{80}$Cu. Our experimental $P_{\rm n}$ values 
for $^{77}$Cu and $^{78}$Cu confirm the finding of an 
experiment carried out in parallel to this study
\cite{WIR09} that the previously reported $P_{\rm n}$ values
are too small, though in the case of $^{78}$Cu we find 
the discrepancy is somewhat smaller as reported in \cite{WIR09}. 
This discrepancy is not too surprising given the large error bars and the 
difficulties encountered in these earlier measurements. 

Given these discrepancies with earlier work, our measurement 
of  the $P_{\rm n}$ value of $^{79}$Cu is of particular importance
because of the key role this nucleus plays in r-process models. 
We find that in this case our new measurement agrees well 
with work reported earlier indicating that the problems with earlier 
experiments are not necessarily of general nature. 

Our experimental data provide constraints for theoretical 
models used to understand the nuclear structure of the 
$^{78}$Ni region. 
We find that recent shell model calculations as well 
as local continuum QRPA calculations describe the experimental
data  well. This gives some confidence that such models can be 
used to investigate nuclear structure effects around 
$^{78}$Ni. However, in the case of the shell model calculations
the required large GT quenching factor indicates the need for larger 
model spaces. 

Our results also show that the global QRPA03 model used to predict 
nuclear structure input for astrophysical r-process 
calculations agrees with data within the expected 
theoretical uncertainty. The model exhibits some local deficiencies in the 
$^{78}$Ni mass region, in particular for the $P_{\rm n}$ values
of the neutron rich Cu isotopes. More experimental
data are needed to identify the nature of this problem. With our measurements 
experimental decay data are now available for r-process
calculations along the entire reaction path at $N=50$. This 
includes our improved data on the branchings for $\beta$-delayed 
neutron emission of $^{78}$Cu and $^{79}$Cu, which 
are needed in r-process model calculations to 
reliably calculate the final abundances in the $A=78-79$
mass range. 
Together with recent mass measurements around 
$^{80}$Zn this now puts a
three isotope section of the r-process around $A=80$ on a fairly solid
experimental basis.
Our r-process model calculations
demonstrate that this narrow mass region from $A=78-80$ turns 
out to be well reproduced with a neutron capture flow based
r-process. More experimental nuclear physics 
data beyond $A=80$ are needed to broaden
the mass range in which r-process models can be tested
reliably, but for now neutron captures under typical r-process
conditions cannot be excluded 
as a mechanism for the origin of the elements around $A=80$
not made by the s-process. 

We also show that the $^{78}$Ni half-life does
play a role in HEW r-process models. 
On the other hand, compared to the classical model, 
the HEW model seems to be less sensitive 
to $P_n$ values in the 
Ni-Cu region, at least for our choice of parameters.
We demonstrated 
that the $P_n$ values of 
heavier nuclei do play a critical role in our HEW r-process model.  

\section{Acknowledgements}
This work was supported by NSF grants PHY 08-22648 (Joint Institute for Nuclear 
Astrophysics), PHY 06-06007 (NSCL), and PHY 02-16783, by the Deutsche Forschungsgemeinschaft
(DFG) under contract KR 806/13, and by the Helmholtz Gemeinschaft under grant
VH-VI-061 (VISTARS).


\begin{thebibliography}{9}

\bibitem{BBFH} E.M. Burbidge et al., Rev. of Mod. Phys. \textbf{29}, 547, (1957). 
\bibitem{CTT91} J.J. Cowan, F.-K. Thielemann, and J.W. Truran,
{\it Phys. Rep.}, , \textbf{208}, 267 (1991).
\bibitem{AGT07} M. Arnould, S. Goriely, and K. Takahashi, Phys. Rep. \textbf{450}, 97 (2007).
\bibitem{WWH92} S.E. Woosley and R.D. Hoffman \emph{ Ap. J.}, \textbf{395}, 202 (1992).
\bibitem{TWJ94} K. Takahashi, J. Witti and H.-Th. Janka, \emph{ A\&A} , \textbf{286}, 857 (1994).
\bibitem{BLD06} A. Burrows et al. NewAR. \textbf{50}, 487  (2006).
\bibitem{Cam01} A.G.W. Cameron, \emph{ Ap. J.} , \textbf{562}, 456 (2001).
\bibitem{FFA06} C.L. Fryer et al., Ap. J., \textbf{646}, L131 (2006).
\bibitem{STM01} K. Sumiyoshi et al.Ap. J. \textbf{562}, 880 (2001).
\bibitem{WTI03} S. Wanajo et al. Ap. J. \textbf{593}, 968 (2003).
\bibitem{RLT99} S. Rosswog et al. \emph{ Astron. Astr.} , \textbf{341},499 (1999).
\bibitem{PWH03} J. Pruet, S.E. Woosley, and R.D. Hoffman, Ap. J. \textbf{586}, 1254 (2003).
\bibitem{SuM05} R. Surman and G. C. McLaughlin, \emph{Ap. J.}, \textbf{618}, 397 (2005). 
\bibitem{JMO06} P. Jaikumar, B.S. Meyer, K. Otsuki, and R, Ouyed, nucl-th/0610013.
\bibitem{SCG08} C. Sneden, J.J. Cowan, and R. Gallino, Ann. Rev. Astron. Astrophys. \textbf{46}, 
241 (2008).
\bibitem{KBT93} K.-L. Kratz et al. \emph{ Ap. J.} , \textbf{403}, 216 (1993).
\bibitem{Pfe01} B. Pfeiffer et al., Nucl. Phys. \textbf{A693}, 282 (2001).
\bibitem{Kra07} K.-L. Kratz et al., Ap. J. \textbf{662}, 39 (2007).
\bibitem{KPT00} K.-L. Kratz et al. \emph{ Hyperf. Int.} , \textbf{129}, 185 (2000).
\bibitem{SBF02} J. Shergur et al. \emph{ Phys. Rev. C} , \textbf{65}, 034313 (2002).
\bibitem{Pfeiffer02}
B.Pfeiffer {\it et al.}, Prog. Nucl. Energy {\bf 41}, 39 (2002).
\bibitem{SDA03} O. Sorlin et al. \emph{ Nucl. Phys. A} , \textbf{719}, C193 (2003).
\bibitem{DKW03} I. Dillmann et al. \emph{ Phys. Rev. Lett.} , \textbf{91}, 162503 (2003).
\bibitem{KPA05} K.-L. Kratz et al., Eur. Phys. J. \textbf{A25}, 633 (2005).
\bibitem{Hos05} P. Hosmer et al. \emph{Phys. Rev. Lett.}, \textbf{94}, 112501 (2005). 
\bibitem{MEH06} F. Montes et al., Phys. Rev. C \textbf{73}, 035801 (2006).
\bibitem{AHH09} O. Arndt, Act. Phys. Pol. \textbf{B3}, 437 (2009).
\bibitem{PHA09} J. Pereira et al., Phys. Rev. C \textbf{79}, 035806 (2009).
\bibitem{TSK01} T. Terasawa {\it et al.}, Ap. J. {\bf 562}, 470 (2001).
\bibitem{Kra84} K.-L. Kratz, Nucl. Phys. \textbf{A417}, 447 (1984).
\bibitem{GBB98} R. Grzywacz et al. Phys. Rev. Lett. \textbf{81}, 766 (1998).
\bibitem{MBF00} W.F. Mueller et al. Phys. Rev. C \textbf{61}, 054308 (2000).
\bibitem{FHK01} S. Franchoo et al. Phys. Rev. C \textbf{64}, 054308 (2001).
\bibitem{SGM03} M. Sawicka et al. Phys. Rev. C \textbf{68}, 044304 (2003).
\bibitem{SMG04} M. Sawicka et al. Eur. Phys. J. A \textbf{22}, 455 (2004).
\bibitem{SPG04} M. Sawicka et al. Eur. Phys. J. A \textbf{20}, 109 (2004).
\bibitem{MGB05} C. Mazzocchi et al. Phys. Lett. B \textbf{622}, 45 (2005).
\bibitem{VDG05} J. Van Roosbroeck, Phys. Rev. C {\bf 71}, 054307 (2005). 
\bibitem{EBD99} J. Engel {\it et. al.} Phys. Rev. C {\bf 60}, 014302 (1999).
\bibitem{DGL00} J.M. Daugas et al.Phys. Lett. B \textbf{476}, 213 (2000).
\bibitem{GGF02} H. Grawe et al. Nucl. Phys. A \textbf{704}, 211c (2002).
\bibitem{MPK03}
P. M{\"o}ller, B. Pfeiffer, and K.-L. Kratz, Phys. Rev. C {\bf 67}, 055802 (2003)
\url{http://t16web.lanl.gov/Moller/publications/rspeed2002.html}. 
\bibitem{Lam03}
K. Langanke and G. Martinez-Pinedo Rev. Mod. Phys. {\bf 75}, 819 (2003).
\bibitem{SDV04} N.A. Smirnova et al. Phys. Rev. C {\bf 69}, 044306 (2004).
\bibitem{Bor05} I. N. Borzov, Phys. Rev. C {\bf 71}, 065801 (2005).
\bibitem{LBH04} A.F. Lisetskiy, B.A. Brown, M. Horoi, and H. Grawe, Phys. Rev. C {\bf 70}, 044314 (2004).
\bibitem{LBH05} A.F. Lisetskiy, B.A. Brown, and M. Horoi Eur. Phys. J. A \textbf{25}, 95 (2005).
\bibitem{MSS03} D. J. Morrissey et al. \emph{Nucl. Instr. and Meth. B}, \textbf{204}, 90 (2003). 
\bibitem{Pri03}
J.I.~Prisciandaro, A.C.~Morton, and P.F.~Mantica, Nucl. Instr. Meth. A \textbf{505}, 140 (2003).
\bibitem{Per09} J. Pereira et al., Nucl. Instr. Meth. A \textbf{618}, 275 (2010).
\bibitem{nubase03}
G. Audi {\it et. al.} Nucl. Phys. A {\bf 729}, 3 (2003).
\bibitem{KBC05} U. K{\"o}ster et al., Proceedings of the 3rd International 
Workshop on Nuclear Fission and Fission-Product Spectroscopy, 
AIP Conf. Proc. {\bf 798}, 315 (2005).  
\bibitem{MGB05a} C. Mazzocchi et al. Eur. Phys. J. A \textbf{25}, 93 (2005). 
\bibitem{KSO82} K.-L. Kratz et al. Z. Phys. A \textbf{306}, 239 (1982).
\bibitem{EFH86} B. Ekstr{\"o}m et al. Phys. Scr. \textbf{34}, 614 (1986).
\bibitem{KGM91} K.-L. Kratz et al. Z. Phys. A \textbf{340}, 419 (1991).
\bibitem{WIR09} J. Winger et al. Phys. Rev. Lett. \textbf{102}, 142502 (2009).
\bibitem{MNK97}
P. M{\"o}ller, J. R. Nix, and K.-L. Kratz, Atomic Data and Nucl. Data Tab. {\bf 66}, 131 (1997).
\bibitem{LB08} B.A. Brown and A. F. Lisetskiy, private communication. 
\bibitem{LBS06} A. F. Lisetskiy, B. A. Brown, and H. Schatz,
Proceedings of the 12th International Symposium CAPTURE GAMMA-RAY SPECTROSCOPY AND RELATED TOPICS
Notre Dame, Indiana (USA), 4-9 September 2005, AIP Conf. Proc.{\bf 819}, 483 (2006).
\bibitem{krumlinde84:a} J. Krumlinde and P. M{\"{o}ller},
Nucl.\ Phys.\ {\bf A417}, 419  (1984).
\bibitem{moller90:a}
P. M{\"{o}}ller and J. Randrup,
Nucl.\ Phys.\ {\bf A514}, 1 (1990).
\bibitem{moller95:b}
P. M{\"{o}}ller, J. R. Nix,
W. D. Myers, and W. J. Swiatecki,
Atomic Data Nucl. Data Tables {\bf 59}, 185 (1995).
\bibitem{audi89} G. Audi,
Midstream atomic mass evaluation, private communication (1989),
with four revisions
\bibitem{takahashi72:a}
K. Takahashi, 
Prog.\ Theor.\ Phys.\ {\bf 47}, 1500 (1972).
\bibitem{takahashi73:a}
K. Takahashi, M. Yamada, and T. Kondoh,
Atomic Data Nucl. Data Tables {\bf 12}, 101 (1973).
\bibitem{AME95} G. Audi and A.H. Wapstra, Nucl. Phys. A \textbf{595}, 409 (1995).
\bibitem{Kra06} Kratz, K.-L., Proceedings of the 12th International Symposium CAPTURE GAMMA-RAY SPECTROSCOPY AND RELATED TOPICS
Notre Dame, Indiana (USA), 4-9 September 2005, AIP Conf. Proc.{\bf 819}, 409 (2006). 
\bibitem{moller07:c}
P. M{\"{o}}ller, R. Bengtsson, K.-L. Kratz, and H. Sagawa,  
Proc. International Conference on Nuclear Data and Technology,
April 22--27, 2007, Nice, France, (EDP Sciences, (2008) p. 69, ISBN 978-2-7598-0090-2), and
\url{http://t16web.lanl.gov/Moller/publications/nd2007.html}
\bibitem{BAB08} S. Baruah et al. Phys. Rev. Lett. \textbf{101}, 262501 (2008).
\bibitem{Hak08} J. Hakala et al. Phys. Rev. Lett. \textbf{101}, 052502 (2008).
\bibitem{Flan09} K. Flanagan et al. Phys. Rev. Lett. \textbf{103}, 142501 (2009).
\bibitem{Cow99} J. J. Cowan et al. Ap. J. \textbf{521}, 194 (1999).
\bibitem{AKW99} C. Arlandini et al. Ap. J. \textbf{525}, 886 (1999).
\bibitem{TGA04} C. Travaglio et al. Ap. J. \textbf{601}, 864 (2004).
\bibitem{AME03} A. H. Wapstra, G. Audi, and C. Thibault, Nucl. Phys. A \textbf{729}, 129 (2003).
\bibitem{Wallet07} J. K. Tuli, Nuclear Wallet Cards, 7th Edition (2005). 
\bibitem{KHH90} K.-L. Kratz et al., Z. Phys. A \textbf{336}, 357 (1990). 
\bibitem{MeB97} B. S. Meyer and J. S. Brown, Ap. J. Suppl. \textbf{112}, 199 (1997).
\bibitem{FRR99} C. Freiburghaus et al. Ap. J. \textbf{516}, 318  (1999).
\bibitem{WKM01} S. Wanajo et al., Ap. J. \textbf{554}, 578 (2001).
\bibitem{TBM01} Thompson, T. A., Burrows, A, and Meyer, B.S. Ap. J. \textbf{562}, 887 (2001).
\bibitem{FKM09} K. Farouqi et al., Ap. J. Lett. \textbf{694}, L49 (2009). 
\bibitem{NON-SMOKER} T. Rauscher, F.-K. Thielemann Atomic Data Nuclear Data Tables \textbf{79}, 47 (2001).

\end{thebibliography}
\end{document}